\documentclass[prb,showpacs,amsmath,amssymb,twocolumn]{revtex4-1}
\usepackage[usenames,dvipsnames]{color}
\usepackage{graphicx,textcase}
\usepackage{dcolumn,overpic}

\begin{document}

\title{Exchange parameters of strongly correlated materials: extraction from spin-polarised density functional theory plus dynamical mean field theory}
\author{Y.O. Kvashnin$^1$, O. Gr$\text{\aa}$n$\text{\"a}$s$^1$, I. Di Marco$^1$, M.I. Katsnelson$^{2,3}$, A.I. Lichtenstein$^{3,4}$ and O. Eriksson$^1$}
\affiliation{$^1$ Department of Physics and Astronomy, Division of Materials Theory, Uppsala University, Box 516, SE-75120 Uppsala, Sweden}
\affiliation{$^2$ Radboud University of Nijmegen, Institute for Molecules and Materials, Heijendaalseweg 135, 6525 AJ Nijmegen, The Netherlands}
\affiliation{$^3$ Theoretical Physics and Applied Mathematics Department, Ural Federal University, Mira Str.19,  620002, Ekaterinburg, Russia}
\affiliation{$^4$ Institute of Theoretical Physics, University of Hamburg, Jungiusstrasse 9, 20355 Hamburg, Germany}

\begin{abstract}
In this paper we present an accurate numerical scheme for extracting inter-atomic exchange parameters ($J_{ij}$) of strongly correlated systems, based on first-principles full-potential electronic structure theory. The electronic structure is modelled with the help of a full-potential linear muffin-tin orbital (FP-LMTO) method. The effects of strong electron correlations are considered within the charge self-consistent density functional theory plus dynamical mean-field theory (DFT+DMFT). The exchange parameters are then extracted using the magnetic force theorem, hence all the calculations are performed within a single computational framework. The method allows to investigate how the $J_{ij}$-parameters are affected by dynamical electron correlations. 
In addition to describing the formalism and details of the implementation, we also present magnetic properties of a few commonly discussed systems, characterised by different degrees of electron localisation.
In bcc Fe, treated as a moderately correlated metal, we found a minor renormalisation of the $J_{ij}$ interactions once the dynamical correlations are introduced.
However, generally, if the magnetic coupling has several competing contributions from different orbitals, the redistribution of the spectral weight and changes in the exchange splitting of these states can lead to a dramatic modification of the total interaction parameter.
In NiO we found that both static and dynamical mean-field results provide an adequate description of the exchange interactions, which is somewhat surprising given the fact that these two methods result in quite different electronic structures.
By employing Hubbard-I approximation for the treatment of the $4f$ states in hcp Gd we reproduce the experimentally observed multiplet structure.
The calculated exchange parameters result to be rather close to the ones obtained by treating the $4f$ electrons as non-interacting core states.
\end{abstract}

\maketitle

\section*{Introduction}
The interatomic exchange interaction ($J_{ij}$), together with the magnetic moment, are the key quantities for a microscopic description and understanding of magnetism in real materials.
Being of a purely quantum origin, they define most of the macroscopic properties of magnetic materials, in particular their Curie temperature ($T_c$), the magnon dispersion and the magnetic response to external stimuli such as an applied magnetic field.
Having a sufficiently large $T_c$, preferably above room temperature, is the key ingredient for any magnetic material that is to be used in the new generation electronic devices (see e.g. Ref.~\onlinecite{parkin-nature04}), or for applications in more traditional fields of magnetism such as new permanent magnetic materials.
Therefore, an ability to predict from first principles electronic structure theory the exchange couplings in various materials is an essential step towards achieving important goals in the design of materials for emerging technologies.

From a theoretical viewpoint, the basic microscopic mechanisms which give rise to the exchange couplings are quite well-known.~\cite{fazekas} 
However, when it comes to real systems, predicting the sign and the magnitude of the $J_{ij}$-parameters is a challenging theoretical problem.
In fact the electronic structure of real materials is complex and results into a competition between various exchange mechanisms. Which one happens to dominate is rather difficult to judge \textit{a priori}.

In 1987, a general scheme was proposed for the extraction of the exchange integrals from electronic structure calculations.\cite{lichtenstein-exch}
(The basic idea was published few years earlier in Ref.~\onlinecite{lichtenstein-exch-orig})
Within this approach the energy of the electronic Hamiltonian is mapped onto a classical Heisenberg model of the form:
\begin{equation}
\hat H = - \sum_{i \neq j} J_{ij} \vec{e}_i \cdot \vec{e}_j ,
\label{HH}
\end{equation}
where $\vec e_i$ denotes the unit vector along the magnetic moment at the site $i$.
The $J_{ij}$-parameters between sites $i$ and $j$ are defined as the response to infinitesimally small rotations of the corresponding atomic spins, i.e. in a linear-response manner. 
The approach was successfully applied to a variety of different materials and has become the primary tool for studying inter-site magnetic interactions from first principles.\cite{magnons-jij,DMS-review,ebert-review11,molmag-review06,antropov-nat11,MVV-Mn12-2014}
Note that the exchange parameters defined in this way are, in general, dependent on initial magnetic configuration. 
In itinerant systems such a dependence may be quite strong, which implies the non-Heisenbergian character of the magnetic interactions.\cite{turzhevskii-jij-dpndnce}

In 2000 it was shown that in the presence of dynamical correlations the formalism is still valid if the self-energy $\Sigma$ is local.~\cite{jij-ldapp-2000}
This is exactly the case of the Dynamical Mean-Field Theory (DMFT)\cite{infdim-DMFT}, which, combined with the density functional theory (DFT), is the state-of-the-art method for \textit{ab initio} modelling of the electronic structure of correlated materials.\cite{anisimov-DMFT-1997,AL-MK-LDApp,kotliar-DMFT} 
{In 2006 X. Wan \textit{et al.} used the same formalism to study the prototypical materials with strong electronic correlations, i.e. the transition-metal monoxides.\cite{Savrasov-jijDMFT} 
Despite Wan and coworkers employed several approximations to make the calculations feasible, they managed to successfully compute the exchange parameters, which were then used to simulate the magnon spectra. }
The authors also reported that the best agreement between theory and experiment was achieved by employing a DMFT-type technique. 

In this paper we describe our implementation of the formalism of Ref.~\onlinecite{jij-ldapp-2000} in a full-potential electronic structure method, in which strong electron correlations are treated with DMFT~\cite{rspt1,rspt-book, csc-dmft, patrik-TMO, grechnev-FeCoNi-PRB, igor-prb-tm-09}.
We apply this implementation to several materials with different strength of correlations.
After this Introduction, the present article is organised in the following way. In section I the general computational scheme is presented. Section II is dedicated to the discussion about how the choice of local projections affects the computed exchange parameters. Sections III, IV and V report the results obtained for bcc Fe, NiO, and hcp Gd. The last section will highlight the main conclusions of our study, and perspectives of our future research.

\section{Theoretical methods}
We model the electronic structure with the help of the full-potential linear muffin-tin orbitals (FP-LMTO) code ``RSPt''.\cite{rspt1,rspt-book} 
In order to treat strongly correlated materials, where the conventional DFT fails, we use a combination of DFT and DMFT (DFT+DMFT), as implemented in a FP-LMTO electronic structure method. Details of this implementation have been presented elsewhere~\cite{csc-dmft, patrik-TMO, grechnev-FeCoNi-PRB, igor-prb-tm-09}, and will not be repeated here with an exception of some specific details that will be needed for our discussion. In DFT+DMFT one first selects a subset of electrons that are not described properly by standard local or semi-local exchange-correlation functionals. Then the Kohn-Sham Hamiltonian of the DFT problem is corrected with an explicit two-particle term describing the local Coulomb repulsion $U$ among these localized electrons minus a double counting term. The latter assures that the contributions of added Coulomb term are removed from the Kohn-Sham Hamiltonian, where they are described less accurately (due to the failure of the exchange-correlation functional). The obtained Hamiltonian is an effective Hubbard model and can be solved by DMFT. Once the DMFT cycle is converged, a new electron density is obtained, which leads to a new Kohn-Sham Hamiltonian. All simulations performed in this study are fully converged with respect to the electron density.~\cite{csc-dmft}

The effective impurity problem, which is at the core of the DMFT method, can be solved by means of various techniques. The results presented in Section III, for bcc Fe, were obtained through the Spin-polarized T-matrix plus Fluctuation Exchange (SPTF) solver~\cite{pourovski-sptf}, which is based on perturbation theory. The solver enforces the system to have the Fermi liquid properties and has to be used with care, i.e. only when $U$ is smaller than the bandwidth $W$. This is usually true for the transition metals. The results presented in Section IV for NiO, instead, were obtained by means of the  Exact Diagonalisation (ED) solver. This is one of the most accurate techniques and can be applied for an arbitrary correlation strength. Recently ED has been applied to study the spectral properties of the transition metals monoxides, and lead to a great agreement between theory and experiment.
Finally the results presented in Section VII, for hcp Gd, were obtained in the Hubbard I approximation (HIA),~\cite{AL-MK-LDApp} where the hybridisation between the correlated orbitals and the rest is neglected, hence the impurity electrons can only redistribute themselves within a given $l$-shell. The HIA approximation is usually employed for compounds with extreme localisation of certain valence electrons. For example, it was applied to a series of rare-earth-based compounds and showed a good agreement with to the measured photoemission spectra and ground state properties.~\cite{Patrik-HIA,Lebegue-HIA,Lebeque-TmSe-2005, Litsarev-Ce-prb}

Once the electronic structure problem is solved, one can extract the $J_{ij}$ couplings. 
To this aim we should define a set of local orbitals $|i,\xi,\sigma \rangle$ for the atomic site $i$. In a crystal $i$ indicates both the lattice vector $\vec R$ of the unit cell and the basis vector $\tau$ within the unit cell, so that it points uniquely to a given ion. The exact shape of the local orbitals does not need to be specified now, and we will postpone this topic to the next Section. For now we keep the formulation as general as possible, and we use the global index $\xi$ to refer to a set of quantum numbers labelling the basis functions. $\sigma$ is a spin index, and can have values $\{\uparrow,\downarrow\}$. With this in mind, we can formulate a generalized expression for the inter-site exchange parameters in DFT+DMFT:
\begin{eqnarray}
\label{jij}
J_{ij} = \frac{\text{T}}{4} \sum_{n} \text{Tr} \biggl[ \hat{\Delta}_i (i\omega_n) \hat{G}^{\uparrow}_{ij}(i\omega_n) \hat{\Delta}_j(i\omega_n) \hat{G}^{\downarrow}_{ji}(i\omega_n) \biggl] 
\end{eqnarray}
where the trace is intended over the orbital degrees of freedom $\xi$, T is the temperature, and $\omega_n = 2\pi T (2n+1)$ is the $n$-th fermionic Matsubara frequency.  $\hat{G}^{\sigma}_{ij}$ is the inter-site Green's function between sites $i$ and $j$ and projected over a given spin $\sigma$. We use only one spin index, since in this whole article we assume that contributions due to the spin-orbit coupling are negligible. 
The term $\hat{\Delta}_i$ gives the exchange splitting at the site $i$: 
\begin{eqnarray}
\hat\Delta_i (i\omega_n) = \hat H_{\text{KS}}^{i,\uparrow} + \hat \Sigma_i^{\uparrow}(i\omega_n) - \hat H_{\text{KS}}^{i,\downarrow} - \hat \Sigma_i^{\downarrow}(i\omega_n) ,
\end{eqnarray}
where we introduced the spin- and site-projected Kohn-Sham Hamiltonian $\hat H_{\text{KS}}$ and local self-energy $\hat \Sigma$. The latter comes from the solution of the DMFT equations. In absence of self-energy we recover the standard DFT expression~\cite{lichtenstein-exch}, with a static exchange splitting. In DMFT,\cite{jij-ldapp-2000} instead, due to the fact that the self-energy is dynamical, the exchange field becomes dynamical too. One has to realise that $\hat \Sigma$, in spite of being local, also affects the inter-site Green's function. This is so, because the Green's function is the resolvent of the \textit{entire} Hamiltonian $\hat H$: 
\begin{equation}
\hat G(z) = \frac{1}{z-\hat H}
\end{equation}
Hence the inversion operation mixes all the matrix elements: site on- and off-diagonal ones. From a numerical viewpoint, the computation of Eq.~(2) is not demanding at all, even if the sum runs over an infinite number of Matsubara frequencies. Since each Green's function at high frequencies behaves as $\sim 1/i\omega_n$, the quantity under summation decays at least as fast as $\sim 1/(\omega_n^2)$. In practice the amount of Matsubara points used for evaluation of the self-energy in a standard DMFT cycle (the exact number depends on the temperature T) is more than sufficient for a reliable evaluation of the $J_{ij}$-parameters.

\section{Local orbitals and exchange parameters}
In this section we provide a detailed description of the construction of the localised basis $\{ |i,\xi,\sigma \rangle \}$ for the evaluation of the exchange couplings. Given that also the DFT+DMFT method requires the definition of local orbitals, we will first present our formalism referred to this method. Later on, we will contextualize the discussion to the exchange parameters. The LMTO basis functions\cite{Andersen75} that are used to solve the DFT problem are $ \{| \vec R, \tau, \chi, \sigma \rangle \}$, where $\vec R$ and $\tau$ refer respectively to the lattice site and the basis vector within the unit cell, while $\sigma$ is again the spin. From $ \{| \vec R, \tau, \chi, \sigma \rangle \}$ one can construct the Bloch sums $\{ | \vec k, \tau, \chi, \sigma \rangle \}$ used to solve the DFT eigenvalue problem and subsequently for the DMFT one-electron Green's function. 
In FP-LMTO the index $\chi$ stands for $\{n,l,m,\kappa\}$, where $n$ is the principal quantum number, $l$ the angular quantum number, $m$ the magnetic quantum number, and $\kappa$ the value of the kinetic energy tail, which defines the asymptotic behaviour of the functions at large distances. 
It is possible to use a minimal basis set, using only a single set of $n,l,m$ and $\kappa$. 
However, for a better converged basis, it is useful to adopt a basis with several $\kappa$ values.
This ensures the flexibility of the basis set to describe various shapes of the electron density (see Ref.~\onlinecite{rspt-book} for more details). 

\begin{figure}[t]
\begin{center}
\includegraphics[angle=0,width=60mm]{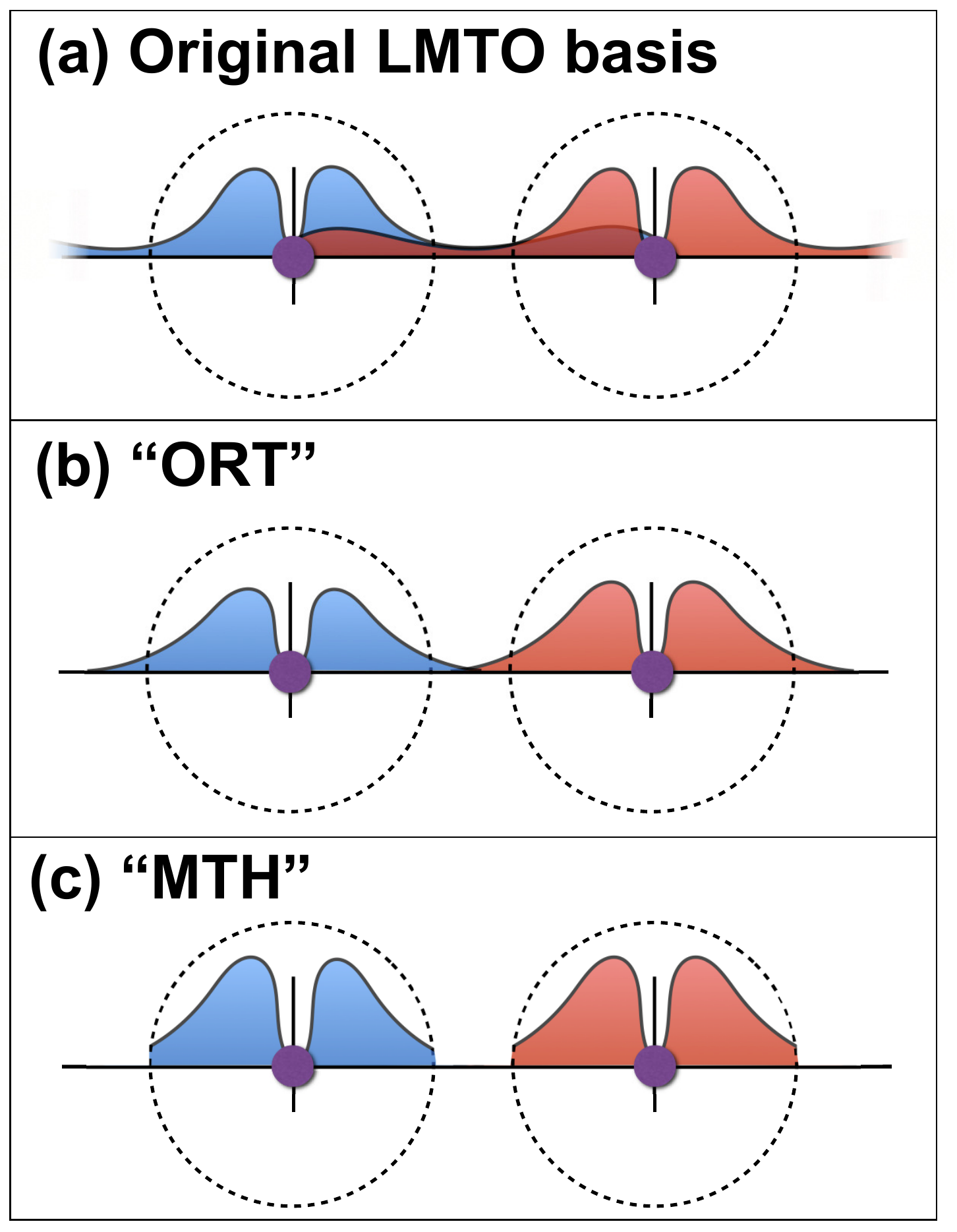}
\caption{Schematic view of the various local orbitals used in RSPt: (a) the non-orthonormal LMTO basis set used in DFT; (b) the L\"owdin orthonormalised LMTO ``ORT'' used as first projection scheme; (c) the muffin-tin head projected orbitals ``MTH'' orbitals used as second projection scheme.}
\label{local_orbitals}
\end{center}
\end{figure}

We consider two types of local orbitals $\{ |i,\xi,\sigma \rangle \}$, which we label as ``ORT'' and ``MTH''. These orbitals are schematically represented in Fig.~\ref{local_orbitals}. The localized orbitals ``ORT'' are constructed directly from the LMTO basis functions $ \{| \vec R, \tau, \chi, \sigma \rangle \}$ (also depicted in Fig.~\ref{local_orbitals}) through the following L\"owdin orthogonalisation:
\begin{eqnarray}
{| i,\xi,\sigma  \rangle}_{\text{\tiny{ORT}}} = \sum_{\vec k, \chi}  e^{i \vec k \vec R_i} \bigl[\hat S_{\vec k, \vec R_0}^{-{1}/{2}}\bigl]_{\chi \sigma, \xi \sigma} | \vec k, \tau_i, \chi, \sigma \rangle 
\end{eqnarray}
where  $S_{\vec k,\vec R_0}$ is the overlap matrix between the LMTO Bloch sums and the local orbitals in the unit-cell at the origin ${\vec R_0}$. Hence, the $\{| i,\xi,\sigma \rangle\}$ are obtained from 
the original basis by performing a $k$-point-wise orthonormalisation. 
In fact, such transformation mixes all the wavefunctions, including the ones centered at different atomic sites. 
Such mixing is anyway small, and the whole construction is analogous to the construction of the Wannier functions.\cite{MLWF-review}
In practice, due to the long decaying tails of the LMTO, these orbitals are not very localized, which is more physical from the electronic point of view. 
Moreover, there is a one-to-one correspondence between the index $\xi$ and the index $\chi$. This may be a problem if one has DFT+DMFT simulations in mind, since an atomic-like basis is required for a proper parametrization of the Coulomb interaction. For the ORT basis this condition is satisfied only when using a single $\kappa$ value. However, multiple $\kappa$ values are often needed for a proper description of the interstitial region between atomic spheres.

The localized orbitals ``MTH'', on the other hand, result from a projection onto the so-called muffin-tin head of the LMTO. The MTH functions are written as:
\begin{eqnarray}
\langle \vec r \mid i,\xi,\sigma \rangle_{\text{\tiny{MTH}}}=
\begin{cases}
 \Phi_{n_\xi l_\xi \sigma}(|\vec r_i|) Y_{l_\xi}^{m_\xi} (\widehat{\vec r_i}), & | \vec r_i | <R_{\text{\tiny{MT}}} \\
    0, & \text{otherwise,} 
\end{cases} \label{eq:MTdef}
\end{eqnarray}
where $r_i = \vec r - \vec R_i$ and $\Phi$ reads as the solution of the radial Schr\"odinger equation inside the MT sphere at the linearization energy~\cite{grechnev-FeCoNi-PRB}. The MTH orbitals are extremely localised, since they are basically atomic functions. They have a pure $l,m$ character and do not overlap by construction. For DFT+DMFT the MTH set is particularly convenient, as in this case $\xi=\{n,l,m\}$, as clear from Eq.~\eqref{eq:MTdef}. There is no need to be in a one-to-one relation with $\chi$, and  one can use basis functions $\{ | \vec k, \tau, \chi, \sigma \rangle \}$ with multiple $\kappa$ for each physical orbital, which results in a very accurate basis. On the other hand, the MTH orbitals completely neglect the interstitial charge and magnetisation densities.

The discussion on the advantages and disadvantages of these projection schemes can be extended to the evaluation of exchange parameters. It is worth to mention that the problem is not entirely the same as that described in the previous paragraphs. For instance, for rare-earth metals, the $4f$ electrons are exposed to pronounced many-body effects, which manifest themselves in the formation of multiplet features in the spectral function. However, it is primarily the $5d$ electrons which participate in the ``communication'' between different atoms by means of exchange interaction. Therefore, the local orbitals to use in DMFT do not have to coincide with those used for the $J_{ij}$ calculation. It was recently shown that the Wannier functions can be employed for the calculations of exchange parameters.\cite{rudenko-jij-wannier} This solution is commonly used in combination with the plane-wave-based DFT codes, since the basis functions are completely itinerant. In LMTO one does not necessarily have to go via this step, because the LMTO basis is atomic-centered and sufficiently localised. The advantage of using the two projections schemes outlined above is that the creation of the basis is less ambigous and perfectly integrated within the electronic structure code. When focusing on the exchange parameters, one can use ORT local orbitals without any restriction on the number of $\kappa$ tails. In fact, when one creates $\{ |i,\xi, \sigma \rangle \}$ uniquely for defining a physical separation of the space, there is nothing that prevents one from using multiple basis functions of the same $l$-character. 

Since the choice of the localised states is a somewhat arbitrary procedure, we should analyse how it affects the final results. The MTH projection totally neglects the interstitial contribution to the magnetisation. This fact already implies that the $J_{ij}$ parameters obtained with different projections will not be the same, simply because the magnetic moments are defined differently. On the other hand, large values of the interstitial magnetisation might be an indication that the system is too itinerant to be studied by means of Eq.~(1). Hence, if the results strongly depend on the projection, the very meaningfulness of the $J_{ij}$ is doubtful in this case.

\begin{figure}[!h]
\includegraphics[angle=0,width=70mm]{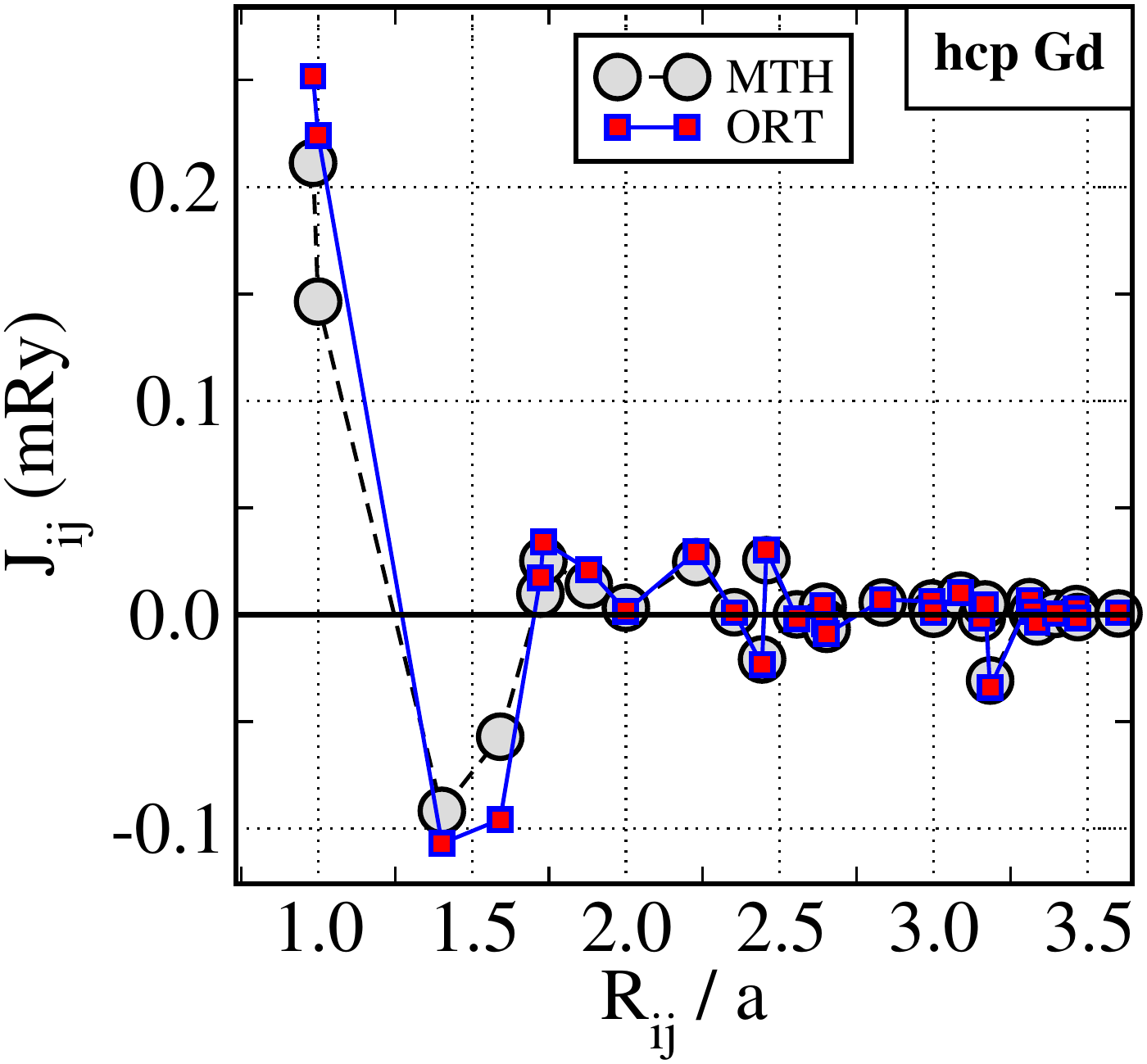}
\includegraphics[angle=0,width=70mm]{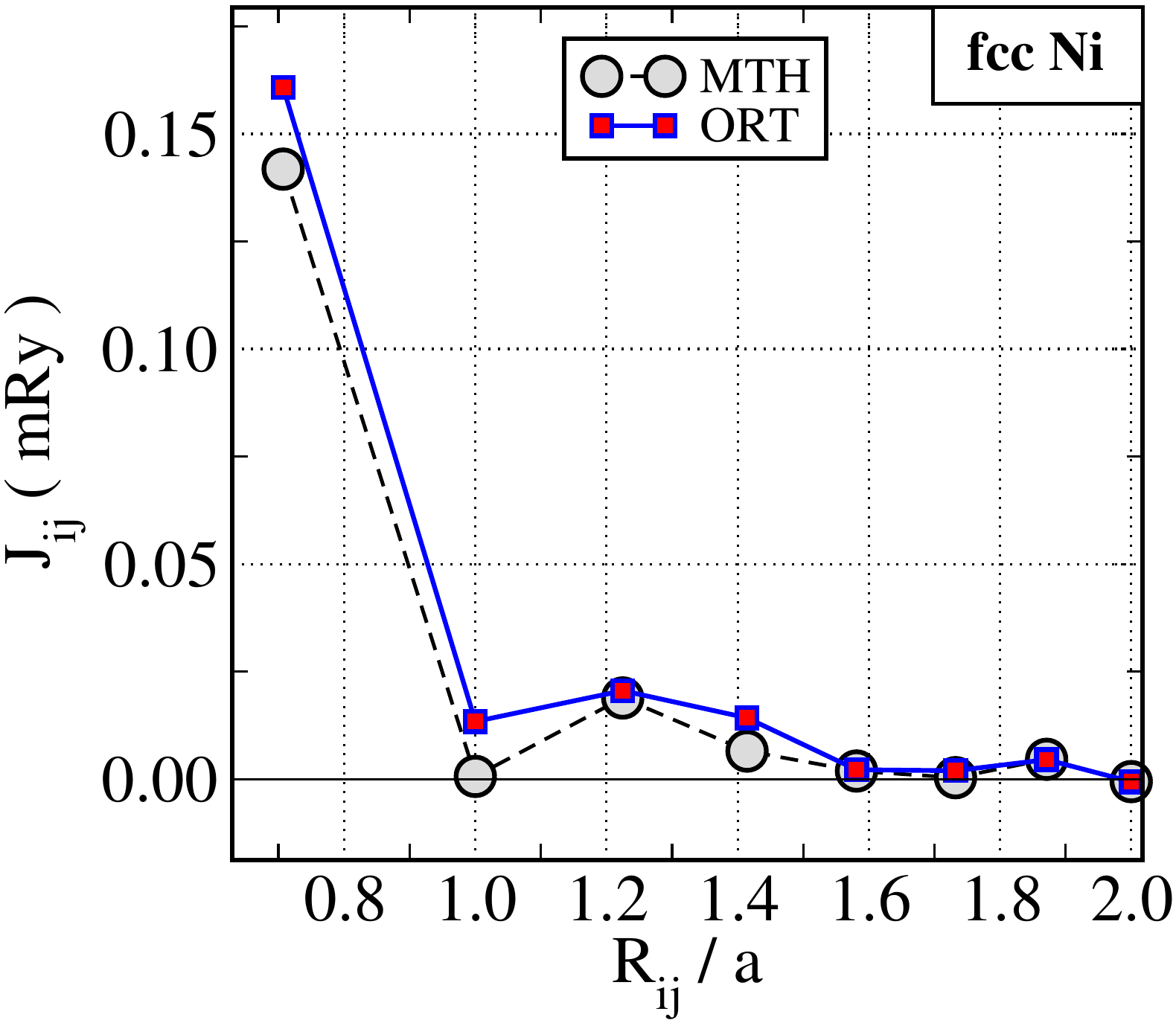}
\caption{Upper panel: inter-site exchange parameters in hcp Gd, extracted from the LSDA, where $4f$ electrons were treated as core. 
Lower panel: inter-site exchange parameters in fcc Ni, extracted from LSDA simulations. 
In each panel two sets of results were obtained by means of the two different projection schemes ORT and MTH. }
\label{gdni-jijs}
\end{figure}

To better illustrate the differences related to the local basis, in the upper panel of Fig.~\ref{gdni-jijs} we report the computed exchange parameters for hcp Gd (upper panel) and fcc Ni (lower panel). The computational setup of Gd is discussed in section IV. The data reported here are obtained from DFT simulations in local spin density approximation (LSDA), and for Gd we treated the $4f$ electrons as non-hybridizing core states.
The magnetic moment carried by the valence electrons inside the muffin-tin sphere is estimated for Gd to be about 0.61 $\mu_B$, in addition to the 7 $\mu_B$ coming from the $4f$ core electrons. Moreover, there is a 0.17 $\mu_B$ per atom of the magnetisation in the interstitial. The latter contributes to the ORT-projected magnetic moment, since the underlying wavefunctions leak out from the MT-spheres (see Fig.~\ref{local_orbitals}, middle panel), but is absent in the MTH case. These differences among the local moments represent the primary reason for the differences between the computed $J_{ij}$-parameters, as seen in Fig.~\ref{gdni-jijs}. Due to the truncation of the wavefunctions, present in MTH projection, the nearest neighbor hopping integrals are expected to be the most renormalised. The same conclusion holds for the exchange couplings, as confirmed by our results. Interactions with more distant neighbours in the two projection schemes are not so different, since the detailed shape of the wavefunctions becomes less important. It is fundamental to note that in the entire range of interatomic distances, the sign and the relative magnitude of interactions is defined unambiguously, i.e. does not depend on the projection scheme.

Excluding the $4f$ electrons from the consideration, magnetisation density in Gd is quite itinerant, which manifests itself in a substantial moment in the interstitial.
Contrary to that situation, magnetisation density in $3d$ metals is much more localised around the atoms.
Hence different projections are supposed to yield very similar results.
For instance, in the lower panel of Fig.~\ref{gdni-jijs} we show calculated $J_{ij}$ parameters in fcc Ni.
We have chosen rather large, almost touching radii of the MT-spheres in order to decrease the size of the interstitial region.
In this case the amount of the magnetisation, which is not attributed to any atom is only about -0.02 $\mu_B$.
Magnetic moment inside the sphere (MTH projection) was computed to be 0.63 $\mu_B$.
An inspection of Fig.~\ref{gdni-jijs} reveals that different definitions of the localised basis sets do not significantly influence the results, even in the present case, where the value of a magnetic moment is relatively small.

\section{Exchange interactions in a weakly correlated metal: bcc iron}
The first system we choose to investigate using the DMFT scheme is bcc iron.
Its electronic structure is well-known and has been studied by means of DFT+DMFT methods by several research groups.\cite{ldapp-Fe-orig,grechnev-FeCoNi-PRB,minar-lsdadmft,igor-prl}
In these studies it was pointed out that iron can be treated as a moderately correlated material. 
Therefore, one can resort to a perturbative solver, such as SPTF with values of $U$ and $J$ parameters equal to 2.3 and 0.9 eV, respectively.
The static part of the self-energy, average over all orbitals per each spin channel, was used as a double-counting correction. With this choice the exchange splitting coming from the DFT part is affected only through the exchange-correlation functional. The temperature was set to T=300 K and 1024 Matsubara points were sufficient to converge the self-energy $\Sigma$ up to the mRy. Both MTH and ORT projections were studied in DFT+DMFT and lead to similar electronic structures~\cite{grechnev-FeCoNi-PRB}. 
For the present study we used two sets of basis functions with different $\kappa$ to describe the $3d$ states, thus MTH projection was employed for the solution of the effective impurity problem. 
After that, the $J_{ij}$ couplings were computed using ORT orbitals, taking advantage of an extended basis set.
However, as discussed above, for transition metals the exchange parameters obtained by means of the two projection schemes are very similar.

\begin{figure}[!h]
\begin{center}
\includegraphics[angle=0,width=70mm]{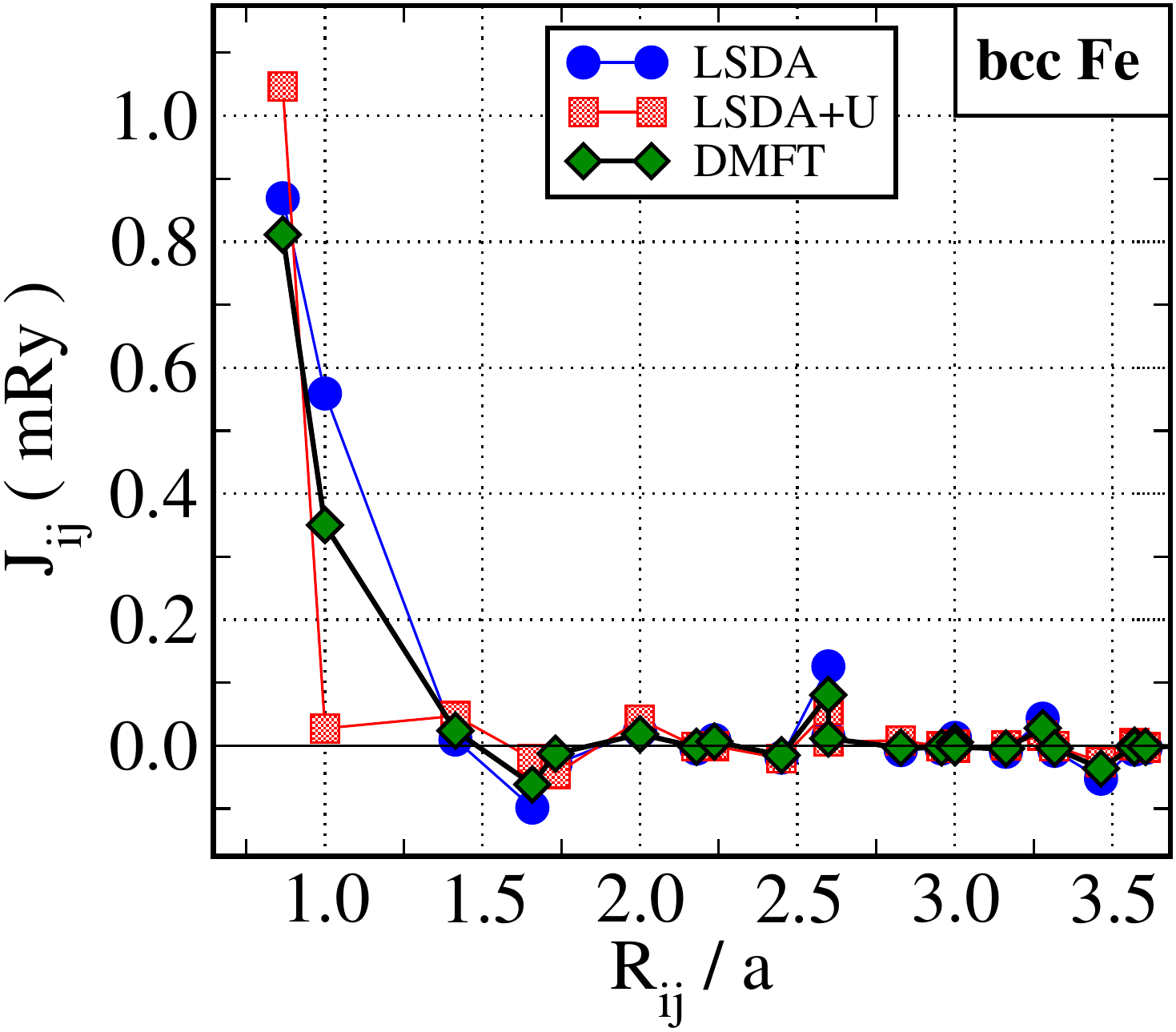}
\caption{Exchange parameters in bcc iron (in mRy) as a function of the interatomic distance (in the units of the lattice constant). Positive sign corresponds to the ferromagnetic coupling. "ORT" projection scheme was employed in the present calculation of the inter-atomic exchange. DMFT results were produced using the SPTF solver.}
\label{Fe-jijs}
\end{center}
\end{figure}

We have extracted the exchange parameters from the converged LSDA, LSDA+U and LSDA+DMFT calculations.
The results of these calculations are shown in Fig.~\ref{Fe-jijs}.
Overall, the obtained $J_{ij}$ couplings are quite similar, at least for the LSDA and LSDA+DMFT computational schemes.
The LSDA+U approach deviates noticeably, at least for the nearest-neighbor (NN) interactions, from the other two methods.

The most drastic difference concerns the 2nd NN interaction ($J_2$), which is significantly suppressed in the case of LSDA+U.
A more detailed analysis of the individual orbital contributions to this exchange parameter reveals drastically different contributions from 3d states with $E_g$ and $T_{2g}$ symmetry, in particular we find that  $J^{E_g-E_g}_2$ is AFM, while $J^{T_{2g}-T_{2g}}_2$ is FM.
This result can also be explained by considering the overlap of the $3d$ wavefunctions.\cite{Bethe-1933}
The lobes of the $T_{2g}$ states of the central atom and its 2nd NN form a 90 degree angle, leading to a relatively small orbital overlap and, therefore, FM coupling. 
The $E_g$ orbitals are pointing towards each other and the exchange interaction is expected to be AFM.
As an example, for LSDA we obtain $J^{E_g-E_g}_2$=-0.06 mRy and $J^{T_{2g}-T_{2g}}_2$= 0.62 mRy. 
Interestingly, in the LSDA+$U$ approximation the local exchange splitting of the $E_g$ states, being about 0.21 Ry, is enhanced with respect to the LSDA value of 0.16 Ry. 
The opposite tendency is found for the $T_{2g}$ orbitals (0.06 Ry versus 0.13 Ry, respectively). 
Since  the local exchange splitting $\Delta$ enters Eq.~(2) twice, any changes of this quantity strongly influence the effective exchange parameter. 
As a result, the $J^{E_g-E_g}_2$ completely compensates the $J^{T_{2g}-T_{2g}}_2$ contribution in the case of LSDA+U. 
We checked that further increase of $U$ leads to the overall change of the $J_2$ sign, because the AFM contribution starts to prevail.

The data shown in Fig.~\ref{Fe-jijs} was produced using the same $U$ and $J$ parameters for LSDA+U and LSDA+DMFT. The latter approach, however, contains also the $3d$ screening on the effective on-site potential, which is totally neglected in the static mean-field LSDA+U.
Therefore, for the same strength of $U$ one can expect that the results of LSDA+DMFT will be somehow closer to that of the conventional LSDA calculation.
This expectation is in line with the results of our numerical simulations. We can also use the computed exchange parameters to evaluate the ordering temperature in the mean-field approximation (MFA): $T^{MFA}_c=\frac{2}{3k_B} J_0$, where $J_0=\sum_j J_{0j}$. 
We obtain a value of 925 K in LSDA, 905 K in LSDA+U and 840 K in LSDA+DMFT.
The experimental value of $T_c$ is about 1045 K, but any agreement in absolute numbers should be regarded as somewhat fortunate. In fact we should mention that, in general, the MFA gives higher values of $T_c$ when compared with Monte-Carlo simulations. This is mainly due to that numerical estimations are based on the assumption that there is no short-ranged order above the critical temperature and there is no temperature dependence of the exchange parameters. However, both approximations are in general rather questionable. In fact recent work suggests that the inter-atomic exchange does depend on the temperature driven magnetic configuration.\cite{Attila-noncol} 

\begin{figure}[!h]
\begin{center}
\includegraphics[angle=0,width=70mm]{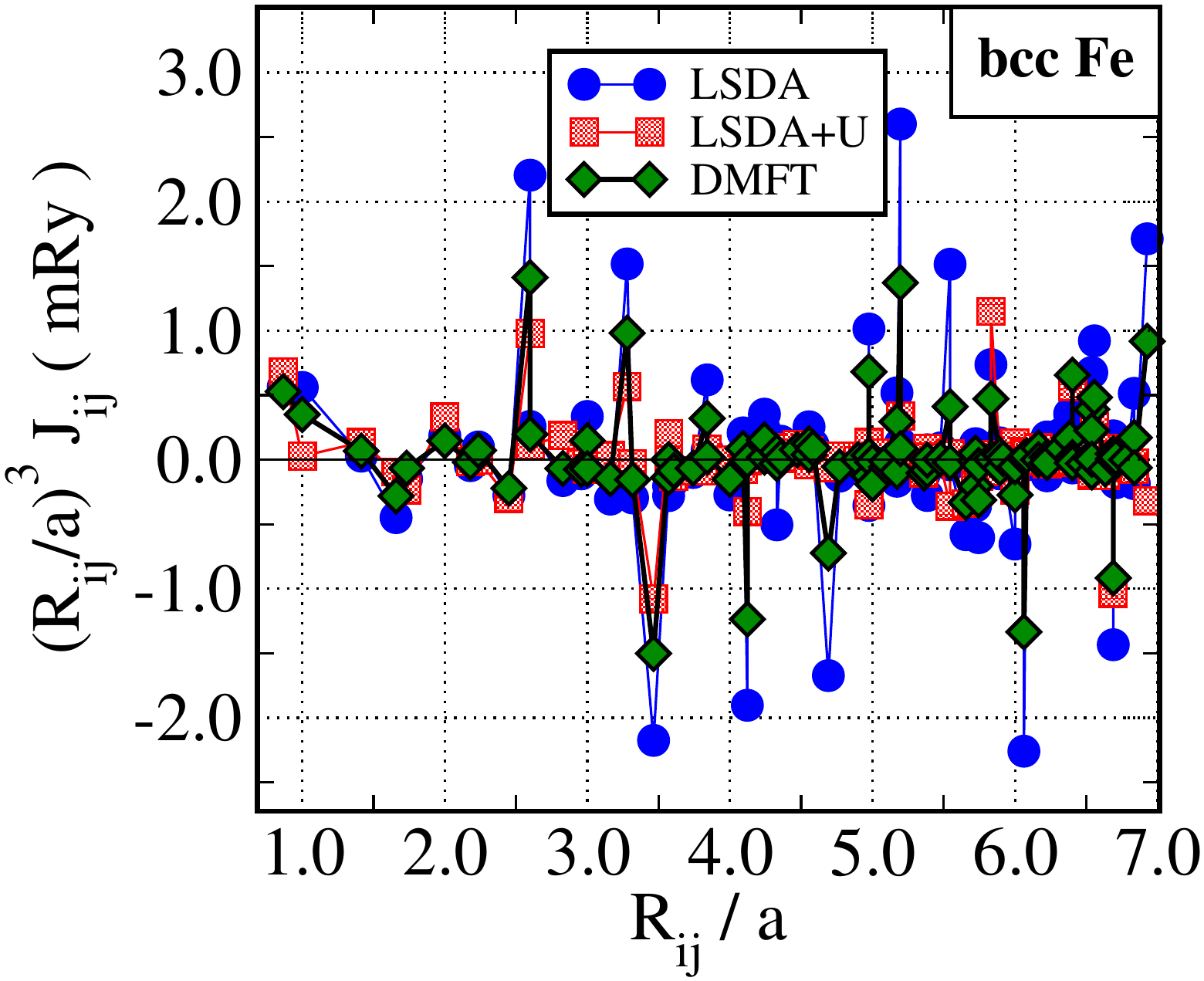}
\caption{Exchange parameters $J_{ij}$ in bcc iron, factorised by inter-atomic distance $R^3_{ij}$.}
\label{Fe-rkky}
\end{center}
\end{figure}

The differences in the interactions for the more distant neighbours, shown in Fig.~\ref{Fe-jijs}, are difficult to distinguish.
It is known that RKKY interactions\cite{rkky} are expected to have a $R^{-3}_{ij}$ asymptotics, where $R_{ij}$ is the interatomic distance.
Hence, in order to illustrate the long-ranged character of the magnetic couplings, Fig.~\ref{Fe-rkky} contains the exchange parameters with the first 99 shells, multiplied with $R_{ij}^3$.
These results were obtained with a very dense k-point grid, consisting of 96$^3$ points in the entire Brillouin zone (BZ).
This is an unnecessary large number for the accuracy of the local quantities, but is, however, necessary for that of the inter-site Green's function.\cite{magnons-jij}

Inspection of Fig.~\ref{Fe-rkky} reveals that indeed LSDA and LSDA+DMFT results are extremely similar.
It is worth mentioning that this is also related to the fact that we use SPTF solver for the DMFT problem, which intrinsically forces moderate correlation effects, but this is what is expected for bulk bcc Fe.
One can see that the effect of dynamical correlations is such that the $J_{ij}$ parameters are renormalised with respect to the LSDA values. 
The renormalisation is, however, anisotropic and is dependent on the direction of the interatomic bond.
Regarding the comparison of the absolute values, one also has to bear in mind that the calculated magnetic moment per Fe is about 2.2 $\mu_B$ in LSDA, whereas it is 1.95 $\mu_B$ and 2.15 $\mu_B$ in LSDA+U and LSDA+DMFT, respectively. 
In the considered form of the Hamiltonian (Eq.~(1)) it means that this difference is effectively contained within the $J_{ij}$ parameters.

\begin{figure}[!h]
\begin{center}
\includegraphics[angle=0,width=70mm]{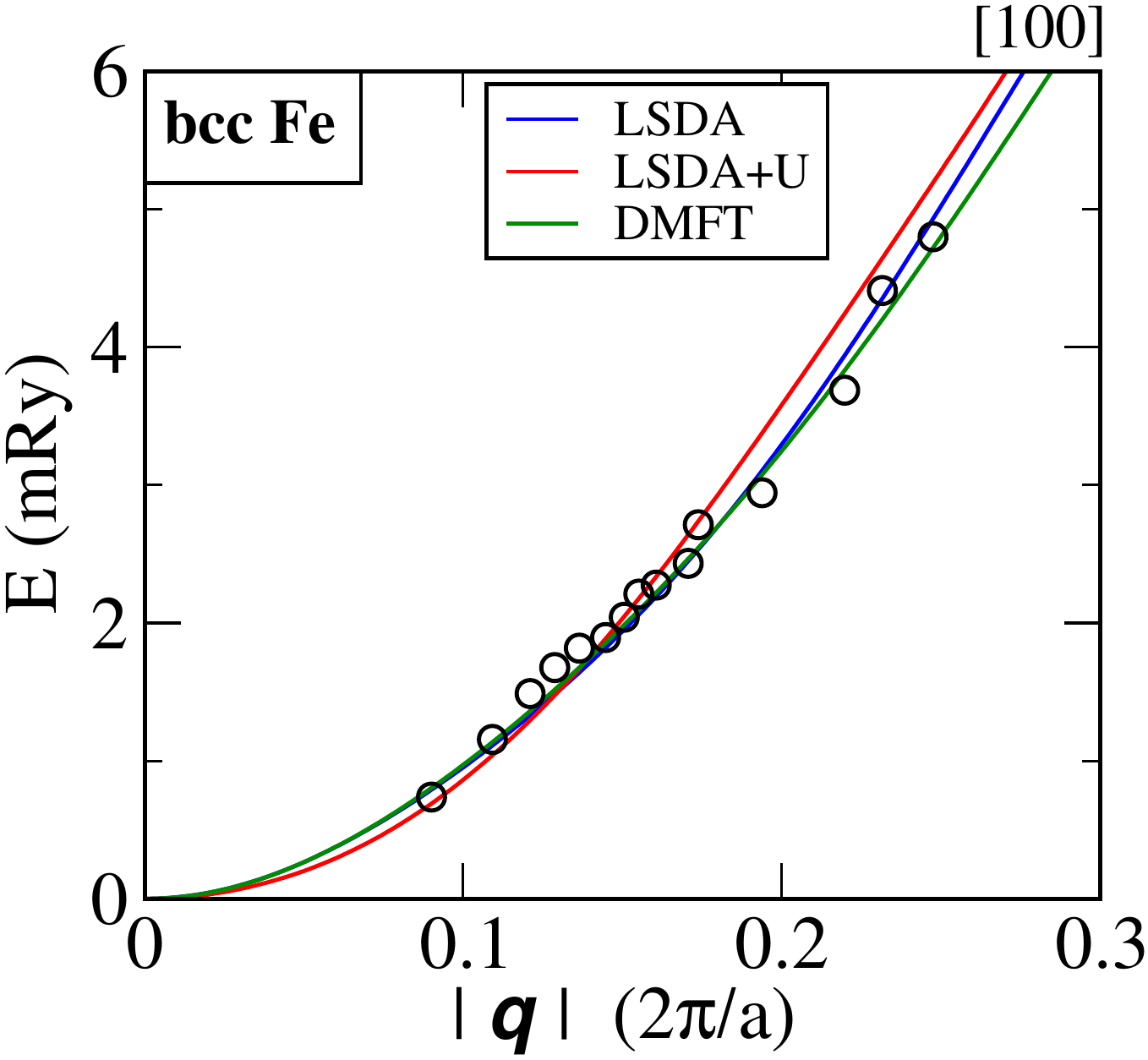}
\caption{Magnon dispersion along [100] direction in the BZ of bcc Fe, simulated using the computed $J_{ij}$-parameters. All calculations were performed for T=300 K. Experiment: room temperature inelastic neutron scattering data (from Ref.~\onlinecite{Fe-RT-magnons}) are shown with circles.}
\label{Esw}
\end{center}
\end{figure}

In order to quantify the differences in the obtained exchange parameters, we have simulated the frozen magnon spectra, expressed through their Fourier transform $J(\vec q)$ (see e.g. Ref.~\onlinecite{magnons-jij}). 
Fig.~\ref{Esw} shows the simulated magnon dispersions along [100] direction together with the inelastic neutron scattering data. 
For small $\mid \vec q\mid$ the results obtained with different electronic structure methods are quite similar. 
Most of the differences become evident at higher momenta, but these regions are difficult to access experimentally. 
Within the considered range of $q$-space, the agreement with experimental data is quite good.

Recently Mazurenko \textit{et al.} have shown that in the Fermi-liquid regime the total exchange interaction $J_0$ for a given site is supposed to be rescaled with an effective mass enhancement factor $Z$.~\cite{mvv-j-rescale}
In general, our results are in line with their conclusion, since we also report an overall decrease in the magnitude of the magnetic couplings once correlations are introduced. 
On the other hand, within the LSDA+DMFT approach the self-energy is different for both spin channels and so is the $Z$ factor. 
Moreover, during the charge self-consistency the unperturbed DFT Hamiltonian gets also modified.
Most importantly, our analysis of the next NN coupling in bcc Fe demonstrated that if there are several exchange paths, the modifications of the exchange parameters can be highly non-trivial.
Thus the relation between $J_0$ obtained in LSDA and that in LSDA+DMFT is more complex.

\section{Strongly correlated system: Nickel oxide}
Next, we have applied our implementation to the canonical example of transition metal oxides (TMOs), namely NiO.
Magnetic interactions in these materials are rather short-ranged, since they are insulating, hence considering a few neighbouring shells is usually sufficient to describe their spin-wave spectra.\cite{MnO-SW, NiO-exp1}. In the literature, the exchange parameters have been successfully described by means of the LSDA+U approach.\cite{Jacobsson-NiO,TMO-Halle,Solovyev-TMOs}
Concerning the details of the electronic structure the LSDA+U is only moderately successful. Magnetic moments and band gap are well described in these materials, but the comparison between experimental photo-electron spectra and theoretical one-particle excitation spectra is far from satisfactory~\cite{kunes-prl-nio}. Hence, in order to reproduce both valence band~\cite{patrik-TMO} and magnon~\cite{Savrasov-jijDMFT} spectra within the same framework one must use the more sophisticated LDA+DMFT method, as we argue here.

Since the correlation effects are known to be very pronounced for these compounds, we could not use the SPTF solver for the impurity problem, but we preferred the ED solver. 
Differently from Ref.~\onlinecite{Savrasov-jijDMFT} we performed DMFT simulations starting from the LSDA solution.
In this way the spin-polarisation already exists at the level of the DFT part of the problem, i.e. before the solution of an effective impurity problem. We employed full-localised limit DC, which is appropriate for the systems close to atomic localisation.~\cite{FLL-DC} 
At convergence, the exchange splitting is due to both the effects of the exchange-correlation functional and those of the dynamical self-energy.\footnote{Contrary to our implementation, the authors of Ref.~\onlinecite{Savrasov-jijDMFT} approximate the exchange splitting to be static at the stage of the evaluation of the $J_{ij}$-parameters.}

The DFT+DMFT simulations with ED were obtained with the same parameters used in Ref.~\onlinecite{patrik-TMO}. The fitting of the hybridization function was done with two bath sites per each $3d$ orbital. The $U$ and $J$ parameters for Ni $3d$ states were set to 8.2 and 0.95 eV, respectively. In order to reach the saturation of the local magnetisation, the temperature was lowered down to 80 K, thus leading to an increase of the number of Matsubara frequencies up to 6144 points. In the following we only present results obtained with the ORT projection.

According to the results of LSDA+DMFT calculation, NiO is characterised by a band gap of about 3.5 eV and 1.85 $\mu_B$ of spin moment per Ni ion. The LSDA+U simulations gave similar results. The calculated magnetic moments are in fair agreement with the experimental value of 1.90 $\mu_B$.\cite{MnO-mom-exp} The calculated exchange parameters in NiO, obtained by means of different techniques are listed in Table I.
\begin{table}[htb]
\centering
\caption [Bset]{Magnetic interactions in NiO, computed by means of various \textit{ab initio} techniques. Exchange parameters $J_N$ (in mRy) denotes the interaction with the $N$-th coordination shell. DMFT results were obtained using ED as an impurity solver.}
\begin {tabular}{c|cc}
  \hline
Computational setup   & J$_1$ & J$_2$  \\
\hline 
 LSDA             & 0.04  & -1.58 \\
 LSDA+DMFT   & -0.003  & -0.48  \\
 LSDA+U         &  -0.002 & -0.50  \\
\hline
LSDA+U (U=8 eV) [Ref.~\onlinecite{Jacobsson-NiO}]    & 0.004 / 0.0 & -0.53  \\
\hline
  Exp.$\#$1 [Ref.~\onlinecite{NiO-exp1}]    & -0.051 & -0.637  \\
  Exp.$\#$2 [Ref.~\onlinecite{NiO-exp2}]    & 0.051 & -0.67  \\
\hline 
\hline
\end {tabular}
\end {table}

It is known that the LSDA gives a rather poor description of the electronic structure of the TMOs. 
In the present calculation we obtained a tiny band gap of about 0.5 eV, much smaller than the experimental estimate.
Since the system is an insulator, at least the underlying exchange mechanism is expected to be correct, i.e. super-exchange is expected.
Note that the situation is much worse for other TMOs, which are predicted to be metals and hence electrons at the Fermi surface participate in the magnetic coupling, resulting in a double exchange mechanism.
As a result, the sign of the leading $J_{ij}$ interaction in NiO is correct, but its absolute value is way too overestimated in LSDA, which is a well-known fact.

\begin{figure}[!h]
\begin{center}
\includegraphics[angle=0,width=80mm]{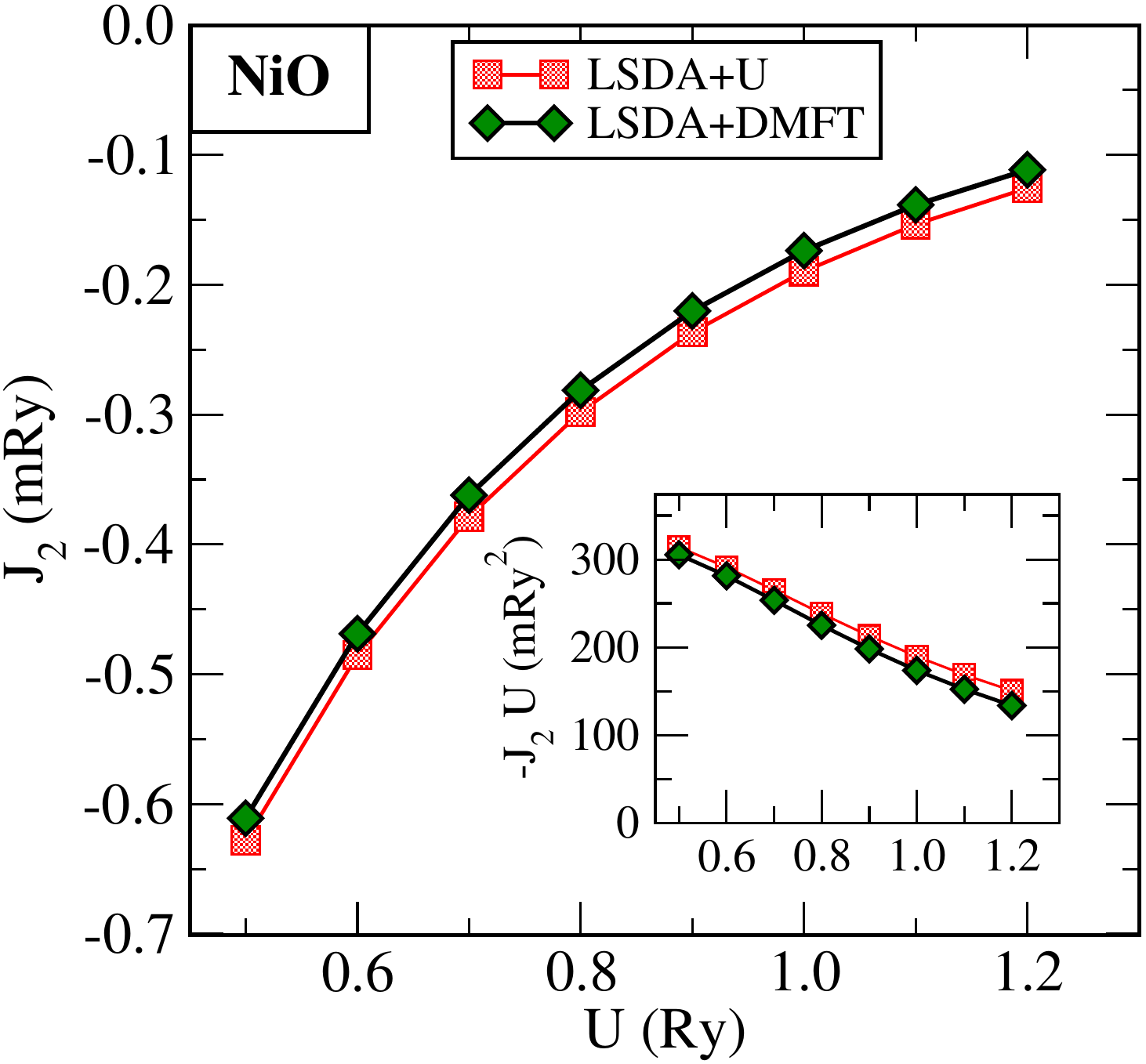}
\caption{Next NN exchange parameter in NiO for various values of Hubbard $U$ calculated within LSDA+U and LSDA+DMFT. Inset: The same plot for $(-J_2*U$).}
\label{t2u}
\end{center}
\end{figure}

Overall, the $J_{ij}$-parameters obtained in DMFT are very similar to those obtained in LSDA+U, both in the present and prior calculations.\cite{Jacobsson-NiO,TMO-Halle,Solovyev-TMOs}
The magnetic interactions in this class of systems are defined primarily by the virtual electron hoppings and relative positions of the electron levels.
Both of these quantities are described relatively well already on the level of LSDA+U.
NiO is a particularly illustrative example, because there is a single $d$ orbital, which participates in the dominant $J_2$ coupling.
The latter, according to the super-exchange mechanism, is expected to be proportional to $t^2/U$, where $t$ is an effective inter-site hopping integral.\cite{pwa-superexch}
To test the relevance of the super-exchange mechanism in a quantitative way, we show in Fig.~\ref{t2u} the calculated $J_2$ coupling as a function of the $U$ parameter.

An inverse proportionality between $J_2$ and $U$ is recognisable. 
If the exchange is given exactly by the term $t^2/U$ one would expect that $J_2*U$ should result in a constant value, $t^2$. 
In order to test this we show $(-J_2*U)$ in the inset of Fig.~\ref{t2u}.
One may note that the so obtained value of $t^2$ is essentially constant and more or less independent on $U$, illustrating that the theory captures correctly the super-exchange mechanism of NiO.
The results of Fig.~\ref{t2u} also imply that the effective hopping parameters extracted from LSDA+U and LSDA+DMFT agree well with each other.
On the other hand, a weak but non-negligible dependence of the effective hopping with respect to $U$ is discernible.
The reason for that is that the calculations are done in the charge self-consistent manner, which implies the change of the DFT potential and, therefore, $t$ values will depend on the $U$ value used in the calculations.
In fact, in these calculations the $U$ value was varied in an unrealistically wide range. 
Considering a narrower window of more reasonable values, one can approximate $t$ to be essentially constant.

Hence we can conclude that the improvements of the use of DMFT are quite moderate for this system, meaning that static local correlations capture most of the essential modifications from the LSDA treatment. 
This is to some extent surprising since the features of the electronic structure of the DMFT and LSDA+U approximations are very different. 
However, one may argue that in order to get the correct value of super-exchange one needs a correct value of the hopping parameter and the gap, and it seems LSDA+U does a good job in providing these properties accurately.
We suspect that the present conclusion holds for most wide gap insulators with large localised magnetic moments.

We also report a fair agreement with experimental data, but a few words about these data are needed. 
From Table I we see that different experiments lead to slightly different values of the $J_{ij}$-parameters, and there is not even consensus about the sign of the NN coupling. 
The experimental data reported in Table I are obtained by fitting the measured magnon spectra to an effective Heisenberg model. 
The precision of this procedure is questionable, especially if one consider that Refs.~\onlinecite{NiO-exp1} and~\onlinecite{NiO-exp2} report on the studies that are forty years old. 
From the numerical side, as seen from Fig.~\ref{t2u}, by choosing a bit smaller $U$ (about 6.8 eV) one can obtain $J_2=0.62$ mRy.
With this respect, we consider our theoretical values in good agreement with experimental data. Our analysis leads to the AFM NN coupling, but the absolute value of this interaction is very small.

\section{Rare-earth metal: hcp G\MakeLowercase{d}}
The $4f$-shell of heavy rare-earth metals is sufficiently localised and essentially does not participate in the chemical bonding.
However, if treated at the LDA level, these states acquire finite dispersion and are placed at the Fermi level in contradiction with existing experimental data.
In particular, in hcp Gd the LSDA approximation leads to an appearance of antiferromagnetic $4f-5d$ exchange interaction, which destabilises the true ferromagnetic ground state (see e.g. Ref.~\onlinecite{OE-Gd-95}). One possibility to overcome this issue is to use the LSDA+U approach, which results in a static $U$-correction, shifting the $4f$ states away from the Fermi level.\cite{FLL-DC} It was also shown that treatment of $4f$ states as purely non-hybridising core levels, improves the description of inter-site magnetic couplings.\cite{Turek-jijs-Gd,Khmelevsky-jijs-Gd} There was, however, a discrepancy in the calculated $J_{ij}$-parameters, which can be explained by the use of different computational codes. Here we have investigated the electronic and magnetic structure of metallic Gd by describing $4f$ states within HIA. The $U$ and $J$ parameters were respectively set to 8.2 eV and 0.7 eV, in agreement with prior studies.\cite{RE-U-abinio,FLL-DC} Given that our localised basis is different from those used in these other studies, a one-to-one correspondence of the exchange parameters can not be expected.

\begin{figure}[!h]
\begin{center}
\includegraphics[angle=0,width=80mm]{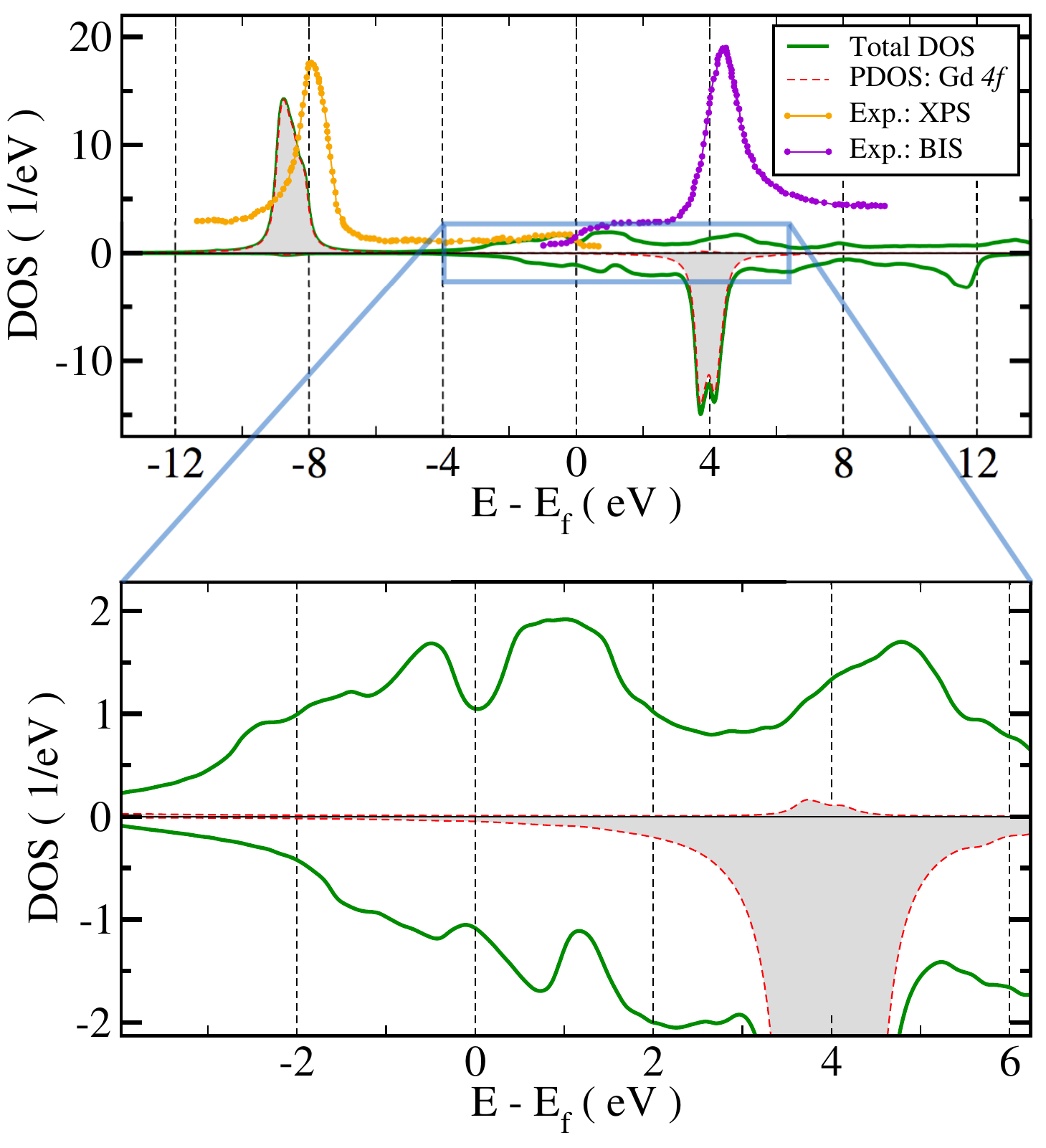}
\caption{Total and $f$-projected spectral function in hcp Gd as computed in LSDA+DMFT method within the Hubbard-I approximation.
The Fermi level is at zero energy. X-ray photoelectron spectroscopy (XPS) and bremsstrahlung isochromat spectroscopy (BIS) data is adapted from Ref.~\onlinecite{exp-rarearths}.}
\label{Gd-rkky}
\end{center}
\end{figure}

The HIA can be seen as a trivial solution in DMFT, where one performs an ED calculation by removing all the hybridization of the local orbitals with all the other states.
However, one is still able to foster multiplet peaks, which are measured experimentally. The calculated electronic structure from the LSDA+DMFT calculations is depicted in Fig.~\ref{Gd-rkky}, where it is compared to experimental data.

An inspection of the plot reveals that the $4f$ states form two distinct groups of multiplets, located around $-8$ eV and 4 eV relative to the Fermi level. 
These two groups of states have purely opposite spin polarisations. 
The agreement with experimental data is acceptable, but can be improved by a slightly different choice of $U$ and $J$, which, however, will not affect the overall picture of the magnetic properties of the system. 
Moreover, the photoemission measurements were done at room temperature, above $T_c$, where the electron levels are spin-degenerate.
We note here that because the ground state of Gd has a filled spin-up shell and empty spin-down shell, the excitations of this configuration become particularly simple and actually quite similar to results from LSDA+U. 
However, for essentially all other rare-earths, either as compounds or in elemental form, the LSDA+U method does not reproduce the measured multiplet structure.

\begin{figure}[!h]
\begin{center}
\includegraphics[angle=0,width=70mm]{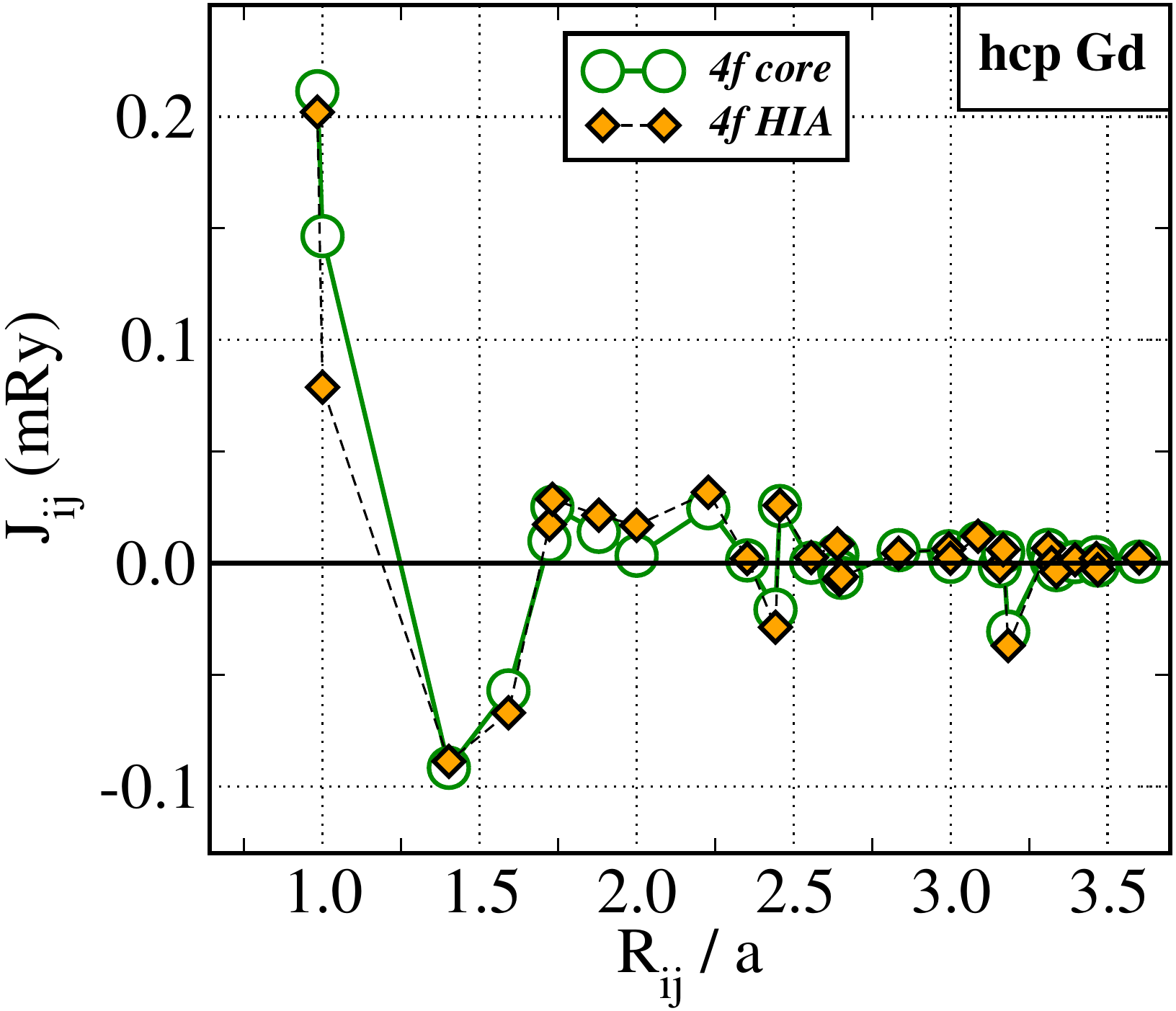}
\caption{Inter-site exchange parameters in hcp Gd, extracted from the ground states with different treatment of $4f$ electrons. The MTH-projection scheme was adopted to produce both sets of results.}
\label{gd-jijs}
\end{center}
\end{figure}

Regarding the exchange interaction parameters, the values are shown in Fig.~\ref{gd-jijs}. 
We found that different treatment of $4f$ states (by means of HIA or simply treating them as part of the core) does not give any qualitative changes.
The most dramatic change is associated with the NN couplings.
There is an anisotropy of the in-plane and out-of-plane exchange parameters, which is a result of an experimentally measured reduced c/a ratio (equal to 1.59 in the current setup).
This difference is enhanced, when the $4f$ electrons enter explicitly in the calculation, which seems to be related to a more non-uniform distribution of the electron density.
The rest of the $J_{ij}$ parameters are quite similar overall: their signs and relative magnitudes are identical.
The spin magnetisation, associated with the most magnetically ``active'' $5d$ electrons, is about 0.48~$\mu_B$. 
This results in an induced valence band moment of 0.68 $\mu_B$/atom (including the interstitial and $6s$/$6p$ MT contributions).
The latter is almost independent on the $4f$ treatment employed.
The computed exchange parameters are of long-ranged nature, clearly showing RKKY oscillations.
Due to this fact, we emphasize that in order to have a converged estimate of $T_c$ or magnon spectrum (not shown here), we have to employ interactions with at least 23 first neighbouring shells. 
Our results are in good agreement with prior studies, given that the computational schemes are different (e.g. an adoption of the atomic sphere approximation).\cite{Turek-jijs-Gd,Khmelevsky-jijs-Gd}
Most importantly, our results are obtained with a full-potential electronic structure code, that can treat open structures, surfaces, etc.

\section{Conclusions}
We report here on the implementation of interatomic exchange parameters for correlated electronic structures and analyse the exchange mechanisms of three classes of correlated electron systems; the weakly correlated transition metal elements, the strongly correlated transition metal oxides and the localized electron systems represented by the rare-earth element Gd. 
We find that details in how electron correlations are included influence the exchange interactions in different amounts, depending on which class one considers. 
To be more specific, we analyse the modifications of exchange couplings in weakly correlated systems, like bcc Fe, as compared to the results of conventional LSDA.
Since the correlation effects act differently on each orbital, their contributions to the total exchange coupling can be substantially different.
In the case of transition metal oxides, static on-site correlations are shown to be the most important ones for an adequate description of the exchange couplings in these systems.
Finally we have applied the method to hcp Gd, where correlated orbitals are of $f$ nature, but more delocalised $spd$ states actually participate in the inter-site magnetic interactions.
We show that the $J_{ij}$-parameters are barely affected by different treatment of $4f$ states, because the corresponding states hybridise weakly with the rest.

Regarding the comparison with available experimental data, it is clear that the LSDA offers an adequate description of the magnetic properties of weakly correlated metals, whereas the LSDA+U is suitable for wide gap insulators. 
However, only the DMFT-based calculations provided us a systematically good agreement for all the materials studied. This fact underlines that the inclusion of electron correlations beyond that of LSDA is in general important.

Using the developed machinery one can study the temperature dependence of the magnetic excitations as one approaches $T_c$.
A few words have to be said about the calculation of the $T_c$. In DMFT one does not take into account all possible finite temperature effects. Even by using the most accurate quantum Monte-Carlo (QMC) solver with an approximate matrix of Coulomb interactions (density-density approximation), there are only longitudinal fluctuations, which contribute to the destabilisation of the long range magnetic order (and hence define the $T_c$). This is opposite with respect to the linear response technique, which we adopt, since here one deals with the transverse spin fluctuations solely. Therefore, the estimates of the $T_c$ obtained with $J_{ij}$'s as calculated here, in combination with MFA or Monte Carlo simulations, are not expected to be the same as the results obtained using only DMFT and QMC.

Recently an extension of the method\cite{lichtenstein-exch,jij-ldapp-2000} was proposed, which permits to extract magnetic couplings from non-collinear spin configurations.\cite{Attila-noncol} 
This is a promising approach, especially if one wants to model the magnetic properties at finite temperature, where the partial reorientation of the spins is present.
Moreover, the employed linear-response approach (at least in the current form) assumes that the dynamics of the spin waves is adiabatic, i.e. the unperturbed electronic structure is used to extract effective exchange parameters.
In general case, the adiabaticity is not guaranteed and magnetic excitations should be regarded as truly out-of-equillibrium processes.
The formulae for calculating the magnon spectra in such case have been recently proposed by Secchi \textit{et al}.\cite{Secchi-noneq-jij}
However, for the moment the method can only be applied to the model systems and its reconciliation with the \textit{ab initio} techniques is far from being at hand.

Many magnetic systems exhibit interesting properties due to the presence of other types of magnetic interactions, originating from the spin-orbit coupling.
Among those there are Dzyaloshinskii-Moriya (DM) interaction\cite{DM-orig1,DM-orig2} and magnetocrystalline anisotropy (MA).
An interplay between different terms gives rise to non-collinear spin structures, observed in multiferroic materials,\cite{TbMnO3-Mostovoy} magnetic ions on surfaces,\cite{Bode-surf-2007} and some low-symmetric TM alloys.\cite{Skyrm-FeCoSi}
Both DM and MA can be calculated using the linear-response-like techniques.\cite{MK-DMI-2009,Solovyev-LaMnO3,Udvardi-DMI-2003,MVV-Mn12-2014,our-nphys-sign-DMI}
However, to our knowledge, the effects of dynamical correlations on the anisotropic parameters were never studied so far.
Our implementation provides an ideal base for conducting an investigation of this problem. 

\section{Ackowledgement}
We authors are grateful to A. Szilva and P. Thunstr\"om for stimulating discussions and assistance in the computations.
A.I.L. acknowledges the financial support from DFG-SFB(668) project.
The work of M.I.K. is supported by the ERC (project 338957 FEMTO/NANO).
O.E. acknowledges the support from VR, the KAW foundation and the ERC.


%

\end{document}